\documentclass[useAMS,usenatbib]{mn2e}
\usepackage{amssymb,amsmath,epsfig,times,psfig,natbib}

\def\yr{{\rm\thinspace yr}}

\def\kpc{{\rm\thinspace kpc}}

\def\kmps{{\rm\thinspace km \thinspace s^{-1}}}

\begin{document}

\title[X-ray variability of NGC1275]{Effects of the variability of the nucleus of NGC1275 on
  X-ray observations of the surrounding intracluster medium}
\author[Fabian et al]{A.C. Fabian$^1$\thanks{Email: 
    acf@ast.cam.ac.uk}, S.A. Walker$^1$, C. Pinto$^1$, H.R. Russell$^1$
  and A.C.Edge$^2$\\
$^1$Institute of Astronomy, Madingley Road, Cambridge CB3 0HA\\
$^2$Department of Physics, Durham University, Durham DH1 3LE}  

\maketitle

\begin{abstract}
The active galaxy NGC1275 lies at the centre of the Perseus cluster of
galaxies, which is the X-ray brightest cluster in the Sky. The nucleus
shows large variability over the past few decades. We compile a
lightcurve of its X-ray emission covering about 40 years and show that
the bright phase around 1980 explains why the inner X-ray bubbles were
not seen in the images taken with the Einstein Observatory. The flux
had dropped considerably by 1992 when images with the ROSAT HRI led to
their discovery. The nucleus is showing a slow X-ray rise since the
first Chandra images in 2000. If it brightens back to the pre-1990
level, then X-ray absorption spectroscopy by ASTRO-H can reveal the
velocity structure of the shocked gas surrounding the inner bubbles.
\end{abstract}


\section{Introduction}

The Perseus cluster of galaxies, the X-ray brightest cluster in the
Sky, has an X-ray cool core surrounding the active central galaxy
NGC1275. The jets forming its radio source 3C84 blow 15~$\kpc$ diameter
bubbles in the hot gas (Boehringer et al 1993; Fabian et al 2000;
2006).  Buoyant ghost bubbles lie further from the nucleus, revealing
the last $10^8 \yr$ or more of energy feedback from the AGN to the cool core.

Radio and $\gamma-$ray light curves of the active nucleus of NGC1275
show large amplitude variability over the past 40 years (Dutson et
al 2014; Abdo et al 2009). The source was bright from the 1960s to about 1990 when it
abruptly  faded by about an order of magnitude. It was faintest around
2000 and now shows signs of increasing again.

Here we look at the long-term variability of NGC1275, finding it to
have dropped by about a factor or 20 between when first imaged with the
Einstein Observatory in 1980 and Chandra in 2000, rising slowly since
then. We demonstrate that the high brightness of the nucleus
prevented detection of the inner bubbles with the Einstein Observatory
in the (relatively) short exposure times used.

We also examine the possibility of detecting resonance line absorption
from the hot gas using the nucleus as a backlight. It appears that if
the the nucleus remains at the 2006 level or similar then such
absorption would be challenging with ASTRO-H, due to its half energy
width of 1.5 arcmin causing the nucleus to appear at low contrast against 
the bright hot gas emission.
It will be straightforward for
Athena with its angular resolution of 5 arcsec or better. If, however,
the nucleus increases to the 1980 level, then it will compromise
emission line studies of the central region with ASTRO-H, but
make absorption spectroscopy possible. Since our line of sight
passes directly through the shocked gas surrounding the $1100\kmps$
 expanding Southern bubble, such studies would give unique insight and
 test the growth and energetics of the bubbles.

\begin{figure*}
  \begin{center}
    \leavevmode
    \hbox{
      \epsfig{figure=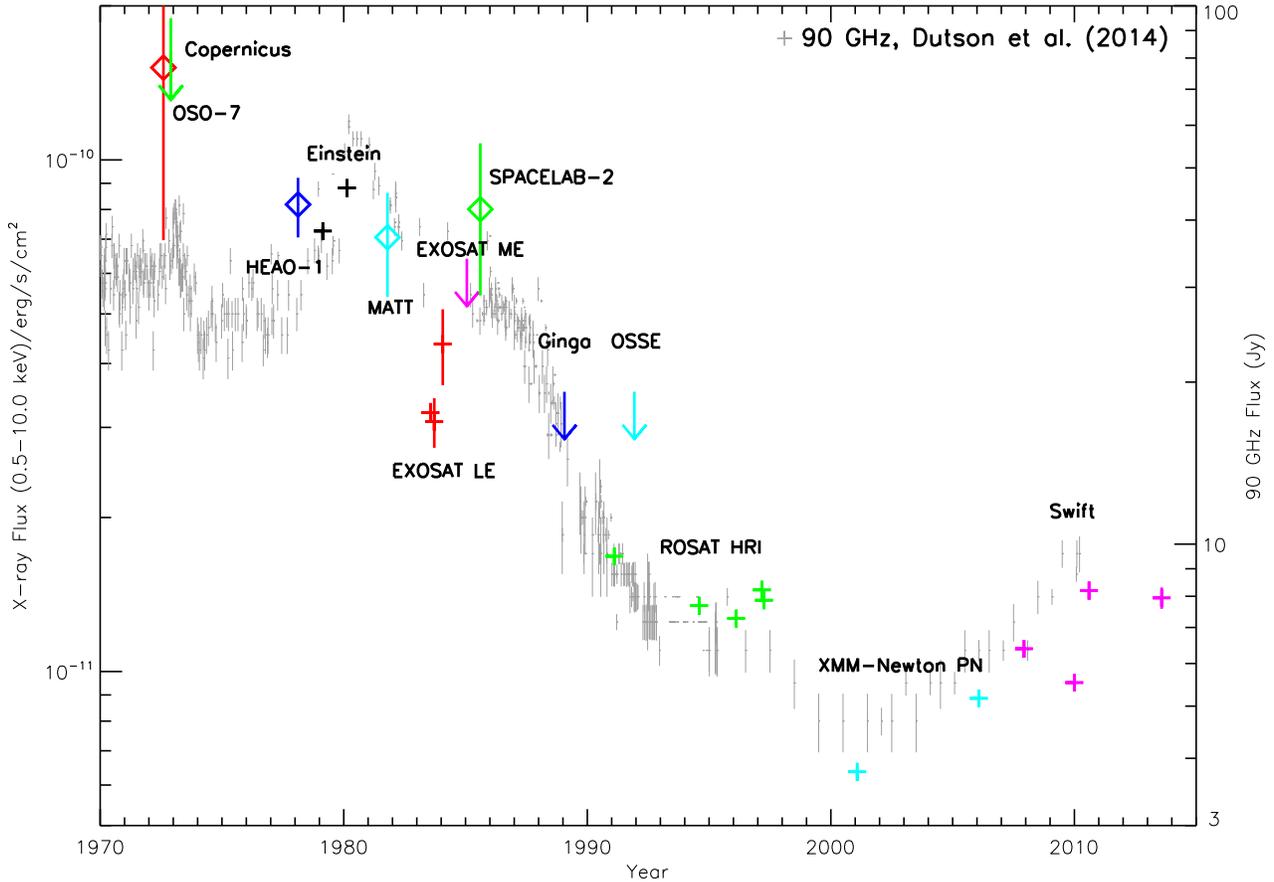,
        width=\linewidth}
            }   
            \caption{Variation in X-ray flux from the central AGN in
              Perseus from the 1970s to the present day. The grey
              points show the 90GHz radio lightcurve gathered by Dutson et
              al. (2014), which follows a similar trend to the X-ray
              lightcurve.}
      \label{Perseus_lightcurve}
  \end{center}
\end{figure*}

\begin{figure*}
  \begin{center}
    \leavevmode
    \hbox{
      \psfig{figure=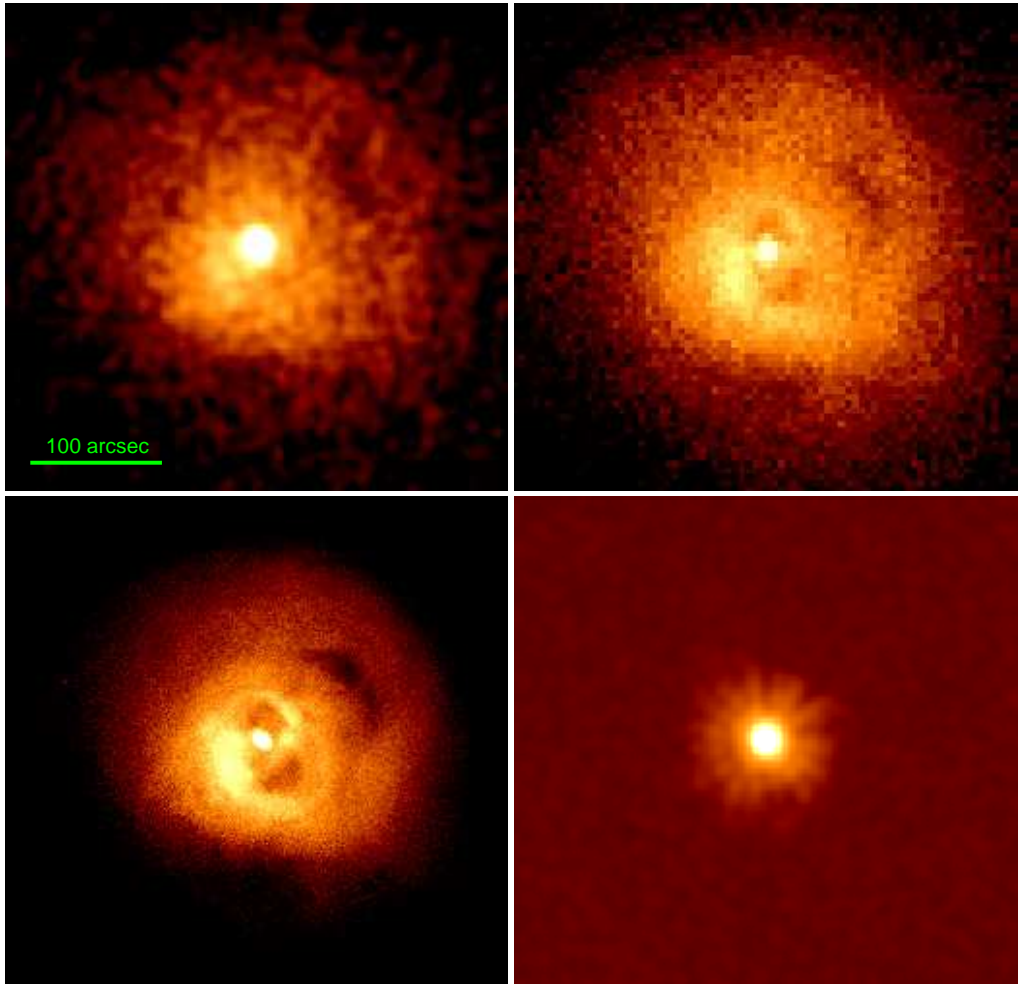,
        width=0.8\linewidth}     
            }
      \caption{Clockwise from top left: Einstein HRI image of NGC1275;
       ROSAT HRI image of NGC1275; Einstein HRI image of the bright
       quasar 3C273 to same angular scale;
       Chandra image of NGC1275. The Einstein image has been lighly
       smoothed. }
      \label{Einstein_imagess}
  \end{center}
\end{figure*}
\begin{table*}
\begin{center}
\caption{Archival Perseus cluster observations }
\label{Archival_obs}

\leavevmode
\begin{tabular}{llllllll} \hline \hline
Telescope & Detector &FWHM & Obs ID & Year & Day  &  Exposure  & Reference \\
           &          & arcsec  &        &          &            &      ks        & \\ \hline
Copernicus &          &120   &        & 1972 Sep &            &                & Fabian et al. (1974)   \\
UCSD OSO-7 &          &  &         & 1972 9-27 Nov  &      &                 & Rothschild et al (1981)\\
HEAO-1 A4 &          &   &         & 1977 Aug; Feb, 1978 Aug &     & &    Primini et al. (1981)\\
Einstein & HRI& 3&  h0316n41.xia  & 1979 &  52 &  15.6 & Fabian et al. (1981)\\
Einstein  & HRI& &  h0316n41.xib  & 1980  & 49 &   6.6 & Fabian et al. (1981)\\
TESRE Balloon &  &       &        &1981 29 Sep  &       &              &  Matt et al. (1990) \\
EXOSAT & LE LX3 & 15 &  ex830724 &  1983  & 205 &   13.4 & Allen et al. (1992) \\
EXOSAT & LE LX3 & &  ex840124  & 1984  & 24 &   2.3 & Allen et al. (1992)\\
SPARTAN-1&     &  &            & 1984 Aug  &     &   &  Ulmer et al. (1987) \\
SPACELAB-2&   &  &             & 1985 Sep  &       & & Eyles et al. (1991) \\
Ginga &      &   &             & 1989 &     15-17       &  &  Allen et al. (1992) \\
OSSE &      &    &             & 1991 28 Nov-12 Dec&   &     &  Osako et al. (1994) \\
ROSAT  & HRI  & 2 & rh800068n00 &  1991 &  41 &   10.9 & Boehringer et al. (1993) \\
ROSAT &  HRI  & &  rh800591a01 &  1994 &  217 &   52.9 &  \\
ROSAT & HRI & & rh702626n00  & 1996  & 42 &   8.2 & \\
ROSAT & HRI & & rh702626a01  & 1997 &  62  &  4.6 & \\
XMM & PN & 7 & P0085110101PN  & 2001  & 30  & 51.2 & Churazov et al. (2003) \\
XMM & PN & & P0305780101PN &  2006  & 29  &  119 &  \\
Swift & XRT& 7 &  sw00036524002xpcw2po  & 2007 &  340 &   3.6 &  \\
Swift & XRT  & & sw00030354003xpcw3po  & 2009 &  364 &   4.3 &  \\
Swift  & XRT & &  sw00031770009xpcw3po &  2010 &  219  &  2.1 &  \\
Swift  & XRT & &  sw00049799006xpcw3po &  2013 &  213  &  1.6 &   \\   \hline

\end{tabular}
\end{center}
\caption{Copernicus:  12, 6, and 2 arcmin apertures;
OSO-7: 6.5 deg FWHM;
HEAO-1: 1.7 x 20 deg;
TESRE balloon: 3 deg;
SPARTAN-1: 5 arcmin x 3 deg;
SPACELAB-2: 12 arcmin;
Ginga: 0.8 x 1.7 deg;
OSSE: 3.8 x 11.4 deg.}
\end{table*} 

\section{Observations}
We investigate the variation in X-ray flux from the central AGN in
Perseus by analysing all of the spatially resolved X-ray observations
that have been taken of the system, starting with the first Einstein
observation from 1979. The observations used are listed in Table
\ref{Archival_obs}.
In addition we have compiled a comprehensive list of hard energy
spectral measurements for NGC 1275 from Copernicus, OSO-7, HEAO-1 A-4, EXOSAT LE,
Ginga LAC, OSSE and several balloon experiments (Table 1). FWHM for
the imaging instruments is tabulated. The early Copernicus estimate contains
an uncertain level of cluster emission so is shown as an upper limit.

From the cleaned events files we find the count rate within a circular
region centred on the AGN of diameter equal to the Full Width at Half
Maximum (FWHM) for the instrument used. The count rate was then
converted into the 0.5-10.0 keV flux using WebPimms, under the
assumption that the spectrum of the AGN is a powerlaw of index 1.65,
as was found in the XMM-Newton study of Churazov et
al. (2003). Residual cluster emission is unlikely to contribute
significantly to the measured nucleus flux when it is bright. 

The resulting X-ray lightcurve is shown in
Fig. \ref{Perseus_lightcurve}, with points from different missions
labelled and shown with different colours. The X-ray lightcurve shows
a decrease by nearly an order of magnitude from the early 1980s to the
1990s. This is in good agreement with the decrease observed in the
90GHz radio emission reported in Dutson et al. (2014), which is
plotted as the grey points.

\section{Simulated Einstein images}
The Einstein Observatory's HRI instrument observed the Perseus cluster
for 15.6ks in 1979 February, and again for 6.6ks in 1980 February. The
central cavities cannot be resolved in either of these images, even
when they are stacked together\footnote{A decrease in X-ray emission at
the position of the the NW ghost cavity is apparent in the Einstein
image, and was commented on at the time (Fabian et al 1981,
Branduardi-Raymont et al 1981). Since it is devoid of detected radio
emission, its true nature was unsuspected.}.
The cavities remained undiscovered
until the first ROSAT HRI observations of Perseus
(Boehringer et al. 1993), which showed for the first time clear
evidence for the interaction between the relativistic particles in the
radio lobes and the intracluster medium (Fig.~2).

\begin{figure*}
  \begin{center}
    \leavevmode
    \hbox{
      \epsfig{figure=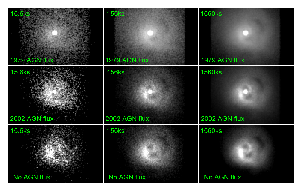,
        width=0.95\linewidth}       
            }
            \caption{Simulated Einstein HRI images with different
              levels of central AGN X-ray flux and different exposure
              times (exposure times are 1$\times$, 10$\times$ and
              100$\times$ the 1979 Einstein observation exposure of
              15.6ks).}
      \label{Einstein_simulations}
  \end{center}
\end{figure*}

The on axis spatial resolution of the Einstein HRI was, however,
comparable to that of the ROSAT HRI (3 arcsec FWHM for Einstein, 2
arcsec FWHM for ROSAT). It is therefore interesting to see whether the
high X-ray flux from the central AGN during the Einstein mission
prevented Einstein from making the first detection of the cavities in
Perseus.

We therefore simulated Einstein images of Perseus to investigate the
visibility of the central cavities as a function of the central AGN
flux and exposure time. The simulations were performed using
\textsc{simx}\footnote{http://hea-www.cfa.harvard.edu/simx/}, which
was modified to allow it to simulate the Einstein observatory. This
was achieved by setting the focal length, field of view and pixel size
parameters equal to those for the Einstein HRI, and using Einstein RMF
and ARF response files. The PSF was set by using the Encircled Energy
Function from Van Speybroeck (1979). For the input image we used
the stacked, 890ks, Chandra image from Fabian et al. (2006). Since only
one spectrum can be used when simulating an events file with
\textsc{simx}, we simulated separate events files for the central AGN
and the cluster emission with different spectra for each, and then
added them. For the central AGN the input spectrum was set as a
powerlaw of index 1.65, as found in Churazov et al. (2003). For the
cluster emission the input spectrum was a thermal \textsc{apec}
component with parameters fixed to the mean cluster temperature in the
core ($kT=$4keV).

We perform Einstein HRI simulations with 3 different AGN flux levels
and 3 different exposure times, and these are shown in
Fig. \ref{Einstein_simulations}. The top row in
Fig. \ref{Einstein_simulations} shows simulations using the 1979 AGN
flux level (7.3$\times$10$^{-11}$ erg s$^{-1}$ cm$^{-2}$ in the
0.5-10keV band) for exposure times of 1$\times$, 10$\times$ and
100$\times$ that of the 15.6ks Einstein observation from 1979. With
the AGN at its 1979 flux level, the PSF spreading prevents the central
cavities from being robustly measured even in very deep (1.5Ms)
observations. The north western `ghost' cavity is however resolved in
the $\geq$156ks simulations.

The second row shows the case when the central AGN flux is reduced to
its minimum observed level (from around 2002) of 5$\times$10$^{-12}$
erg s$^{-1}$ cm$^{-2}$ in the 0.5-10 keV band. We see that a 15.6ks
observation can begin to resolve the inner cavities (1st column, 2nd
row in Fig. \ref{Einstein_simulations}). With ten times this exposure
time, the cavity system is clearly resolved, showing a similar level
of detail as the ROSAT HRI observations.

Finally, in the third row, we show a simulated Einstein image of the
case where there is no contribution from a central AGN (bottom row),
allowing us to demonstrate Einstein's ability to resolve fluctuations
in extended X-ray emission from the ICM in the idealised case of no
point source contamination. We see that even in a 15.6ks observation,
matching the exposure time of the 1979 Einstein observation, the inner
cavities can just about be resolved. Structure is seen on the scale of
the radio source which may at least have triggered further
observation. The cavity system is comfortably resolved with an
exposure time which is ten times deeper.

\section[]{X-RAY ABSORPTION SPECTROSCOPY\\* USING NGC1275}

In this section, we  analyse  absorption by the ICM
acting on the AGN emission, using a multiphase
collisionally-ionized absorber. 
To calculate the column density of gas in each temperature range, we
extracted azimuthally averaged temperature and density profiles from
the Chandra data, which were deprojected using the DSDEPROJ direct
spectral deprojection method (Sanders \& Fabian 2007; Russell, Sanders
\& Fabian 2008). The reduction of the Chandra data is described in
Sanders \& Fabian (2007). We integrated the deprojected density profile
along the line of sight to find the column density in the following
temperature bands which are tabulated in Table \ref{t:absorption_model};
$<$2, 2-3, 3-4, 4-5 and 5-6 keV.

The lowest two temperature components are associated with the high
pressure (shocked) rims around the bubbles (Fig.~6). The density jump
at the outer edge of the rims is 1.3 and the inferred Mach number is 1.2
(Graham et al 2008).

\begin{figure}
\begin{center}
\includegraphics[width=55mm,angle=-90]{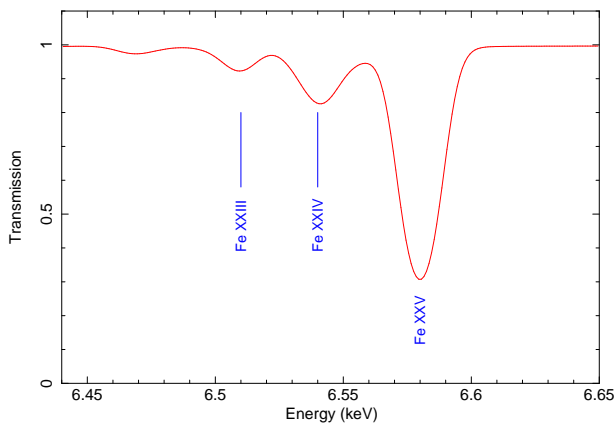}
\end{center}
      \caption{Transmission of the multiphase absorption model.}
          \label{fig:absorption_model}
\vspace{-0.5cm}
\end{figure}

For the absorption, each phase was modeled with a single-temperature
\textit{hot} model in SPEX\footnote{www.sron.nl/spex} (version
2.03.03), which calculates the transmission of a collisionally-ionized
equilibrium plasma. For a given temperature and set of abundances, the
model calculates the ionization balance and then determines all the
ionic column densities by scaling to the prescribed total hydrogen
column density. We used five \textit{hot} components with temperatures
and hydrogen column densities as quoted in
Table\,\ref{t:absorption_model}. We adopted a metallicity of 0.65
(Sanders \& Fabian 2007) and
 a nominal velocity dispersion of 300\,km\,s$^{-1}$
  We show the transmission of
this multi-temperature absorber near the Fe K complex in
Fig.\,\ref{fig:absorption_model} with the strongest absorption line
labeled. An AGN spectrum provided with both high spatial and spectral
resolution would reveal a strong absorption feature at 6.58\,keV.

\begin{table}
\renewcommand{\tabcolsep}{2mm}
\renewcommand{\arraystretch}{1.2}
\footnotesize
\caption{\label{t:absorption_model} Multiphase absorption model}
\begin{center}
\begin{tabular}{lll}
  \hline
  T$_{\rm range}$\,(keV)    &T$_{\rm adopted}$\,(keV)    & $N_{\rm H}  (10^{21}$\,cm$^{-2}$) \\
  \hline                                                                                                                               
  $< 2.0$    & 1.0  & 2.3  \\
  $2.0-3.0$ & 2.5  & 5.0  \\
  $3.0-4.0$ & 3.5  & 4.0  \\
  $4.0-5.0$ & 4.5  & 2.8 \\
  $5.0-6.0$ & 5.5  & 1.0 \\
  \hline
\end{tabular}
\end{center} {\footnotesize Notes: the model consists of five
  isothermal ionized absorber in collisional equilibrium, each of
  which has fixed temperature T$_{\rm adopted}$ and hydrogen column
  density $N_{\rm H}$.}
\end{table}

\begin{figure}
\begin{center}
\includegraphics[width=55mm,angle=-90]{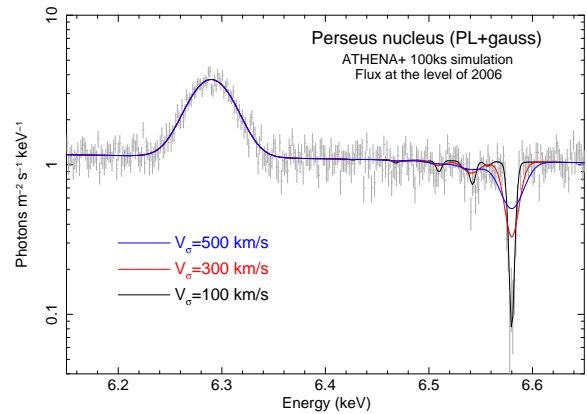}
\end{center}
      \caption{ATHENA/X-IFU simulation for the inner 14" .}
          \label{fig:athena}
\vspace{-0.5cm}
\end{figure}

In order to estimate the nuclear emission of NGC1275, we have analyzed
the longest on-axis exposure of the Perseus cluster taken with
XMM-Newton/EPIC in 2006 (ID=0305780101). Data were reduced and
corrected by solar flares with the SAS v14 following the standard
XMM-SAS procedure. We extracted the nuclear spectrum and subtracted
the cluster spectrum following the procedure of Churazov et
al. (2003). The cluster-subtracted nucleus spectrum was then fitted
with a power law and a gaussian to model the nuclear Fe K emission
line, both redshifted and absorbed by the Galactic foreground. EPIC
spectra do not have spectral resolution high enough to resolve any
narrow absorption lines of the foreground ICM.

The X-ray integral field unit (X-IFU) aboard ATHENA is planned to have 
spatial (5~arcsec) and spectral (2.5eV) resolution at 6\,keV high
enough to resolve the Fe K lines absorbed by the nucleus. We
used the nuclear spectral model fitted with the EPIC spectra as a
template model to simulate a 100\,ks X-IFU spectrum. On top of the AGN
emission the multiphase absorber was tested with three different
values of velocity dispersion $v_{\sigma}$ (see
Fig.\,\ref{fig:athena}). The high-quality X-IFU spectrum will easily 
resolve the lines.

The ASTRO-H Soft X-ray Spectrometer (SXS) with its 5\,eV spectral
resolution could potentially resolve absorption in the nucleus
emission of NGC1275, but due to the large 1.5~arcmin HEW, the SXS
central chip will include both the central AGN and the cluster
emission from a region of about 1.5~arcmin width. In order to
estimate the cluster emission, we extracted a further EPIC spectrum in
an annulus of 0.75~arcmin outer radius and inner radius equal to that
of the nuclear region (14~arcsec, see Churazov et al. 2003). We
modelled the cluster spectrum with a 4\,keV isothermal
collisionally-ionized emission model (CIE model in SPEX). The
abundances of all the atomic species from carbon to iron were fixed to
0.65 similarly to the multiphase absorber.

\begin{figure*}
  \begin{center}
    \leavevmode
    \hbox{
      \psfig{figure=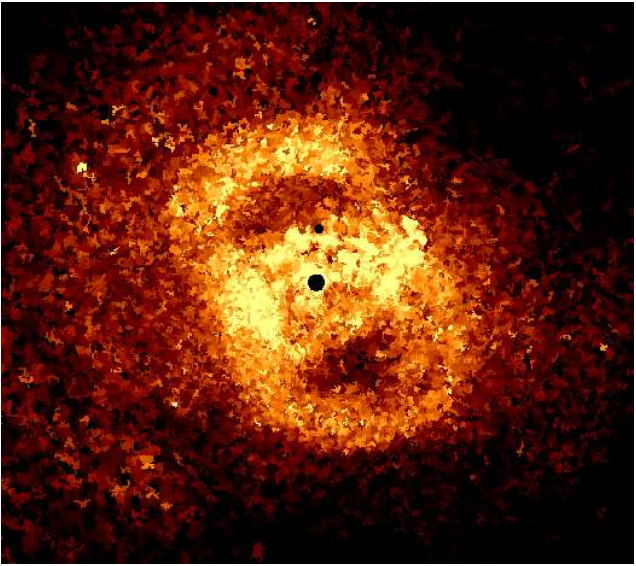,
        width=0.5\linewidth}
      \epsfig{figure=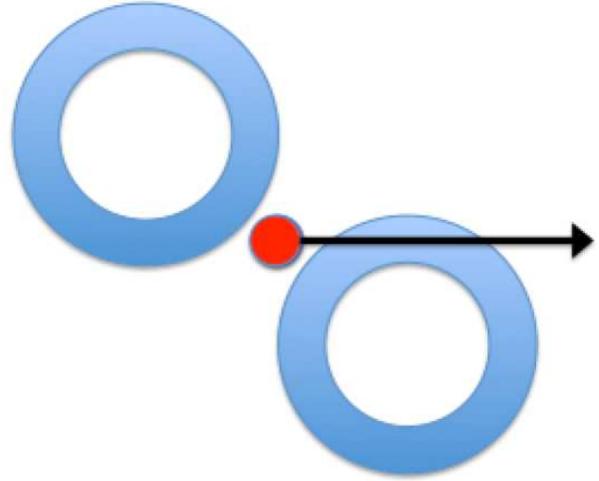,
        width=0.5\linewidth}
            }
      \caption{Left: Pressure map of the region around NGC1275. The
        bubbles are seen to be surrounded by spherical high pressure
        rings of shocked gas (see e.g. Graham et al 2007). Right:
        schematic diagram showing how the bubbles probably lie along
        our line of sight, with the nucleus being viewed through the
        top part of the S bubble.}
      \label{Chandra_Pressure}
  \end{center}
\end{figure*}

\begin{figure}
\begin{center}
  \includegraphics[width=53.5mm,angle=-90]{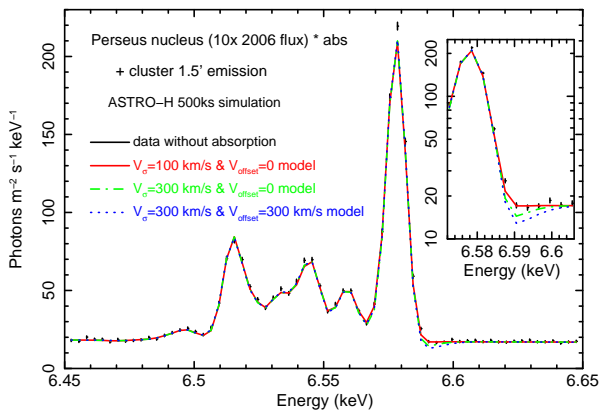}
\end{center}
\caption{\small{ASTRO-H/SXS simulation of the central 1.5' diameter cluster
    emission on top of an AGN spectrum 10x brighter than in
    2006. Three models of AGN instrinsic absorption are shown for a
    combination of different velocity dispersion and line-of-sight
    velocity.}}
          \label{fig:astroh}
\vspace{-0.5cm}
\end{figure}

We use these EPIC spectral fits to finally define a hybrid model
which includes the nuclear emission (power law and Fe K line)
intrinsically absorbed by the multiphase ICM plus the unabsorbed 1.5 arcmin
cluster emission (CIE), both corrected by redshift and Galactic
foreground. We simulate an ASTRO-H/SXS 100\,ks exposure for two models
with and without AGN intrinsic absorption. We could not distinguish
between the models because the cluster accounts for more than $95$\% of
the emission within 1.5 arcmin. A very long exposure above 500\,ks could
potentially reveal some weak evidence of absorption. However, as
previously mentioned, the AGN is brightening in recent years and
at some point it may be as bright as in 1980, which was 10 time
brighter than in 2006. We have therefore simulated a 500ks ASTRO-H/SXS
spectrum with three models of AGN
absorption calculated for a combination of different velocity
dispersion and line-of-sight velocity of the multiphase absorber (see
Fig.\,\ref{fig:astroh}).  The lowest (blue) line in this Figure
includes a velocity offset of $-300\kmps$ (the bubble is likely rising
buoyantly in the cluster potential) with a velocity dispersion
of $300\kmps$ as a rough approximation of the expected velocity
structure behind the weak shock along our line of sight (Fig.~6). 
The effects of absorption of the AGN
are clearly noticeable. 

Resonance line scattering of the emission from the intracluster medium
in the cluster is discussed in detail by Zhuravleva et al (2013) and
is not included in our modelling. We note that the largest red and
blueshifted absorption components seen in the spectrum are likely to
be due to the expanding inner bubbles.

We note that the shape of the 6.4 keV neural iron line (shown in Fig.~5) may
hold interesting information (Churazov et al 1998) about the velocity
structure of the innermost massive molecular clouds surrounding the
nucleus (Salom\'e et al 2006; see also Jaffe 1990 for HI absorption
against the nucleus).

\section{Discussion}

We have shown that the X-ray bright nucleus can play both a negative
and positive role in the detectability of features in the intracluster
medium. When the nucleus of NGC1275 was very bright back in 1980 it
prevented detection of the inner cavities. Our understanding of AGN
feedback in clusters would have begun much earlier had the nucleus
been weak at that time. If the nucleus brightens to a similar level 
in the next decade then it will enable unique absorption spectroscopy of the
high pressure expanding rim around the inner Southern cavity to be
carried out.

Large variability of the jetted nucleus emission is common in cool
core clusters. HST-1, part of the inner jet of M87, brightened and
faded in X-rays by over an order of magnitude between 2004 and 2008
(Harris et al 2009). Russell et al (2013) find significant X-ray
variability in the nucleus of A2052 and Hydra-A; Hogan et al (2015)
find high-frequency radio variability in 4 out of 23 central cluster
sources.  As shown above, an X-ray bright nucleus can enable X-ray
absorption spectroscopy. The quasar H1821+643 at the centre of an
X-ray luminous cluster at redshift $z-0.297$ ( Fang et al 2002;
Russell et al 2010; Walker et al 2014) is a potentially good target.

\section*{Acknowledgements}
ACF, SAW and HRR acknowledge support from ERC AdG FEEDBACK. ACE
acknowledges support from STFC grant ST/I001573/1

\section*{REFERENCES}
\bibliographystyle{mn2e} \bibliography{Perseus_cavities_paper}
Abdo A.A., et al. 2009, ApJ, 699, 31\\
Allen, S. W. et al. 1992, MNRAS, 254, 51 \\
Boehringer H., Voges W., Fabian A. C., Edge A. C., Neumann D.
M., 1993, MNRAS, 264, L25 \\
Branduardi-Raymont G., Fabricant D., Feigelson E., Gorenstein P.,
Grindlay J., Soltan A., Zamorani G., 1981,
ApJ, 248, 55 \\
Churazov E., Sunyaev R., Gilfanov M., Forman W., Jones C., 1998, MNRAS, 297, 1274 \\
Churazov E., Forman W., Jones C., Bohringer H., 2003, ApJ, 590,225 \\
Dutson, K. L., Edge, A. C., Hinton, J. A., et al. 2014, MNRAS, 442, 2048 \\
Eyles C., et al. 1991, ApJ, 376, 23 \\
Fabian, A.C.,et al. 1974 ApJLett 189, L59. \\
Fabian, A. C. et al. 1981, ApJ, 248, 47 \\
Fabian A. C. et al., 2000, MNRAS, 318, L65 \\
Fabian A. C., Sanders J. S., Taylor G. B., Allen S. W., Crawford
C. S., Johnstone R. M., Iwasawa K., 2006, MNRAS, 366, 417 \\
Fang T., et al., 2002, ApJ, 565, 86\\
Graham, J., Fabian, A. C., \& Sanders, J. S. 2008, MNRAS, 386, 278 \\
Harris, D. E., Cheung, C. C., Stawarz, L., Biretta, J. A. \& Perlman,
E. S., 2009, ApJ, 699, 305 \\
Hogan, M., et al., 2015, MNRAS in press\\
Jaffe W., 1990, A\&A, 240, 254\\
Matt, G. et al. 1990 ApJ 255, 468. \\
Osako C.Y., et al. 1994, ApJ, 435, 181 \\
Primini F.A., et al. 1981, ApJLett, 243, L13 \\
Rothschild R., et al (1981) ApJLett 243, L9 \\
Russell H. R., Sanders J. S., Fabian A. C., 2008, MNRAS, 390, 1207 \\
Russell H. R., Fabian A. C., Sanders J. S., Johnstone R. M., Blundell
K. M., Brandt W. N., Crawford C. S., 2010, MNRAS, 402, 1561 \\
Russell, H.R., McNamara, B.R., Edge, A.C., Hogan, M.T., Main, R.A.,
Vantyghem, A. N. 2013, MNRAS, 432, 530\\
Salom\'e, P., Combes, F., Edge, A. C., et al. 2006, A\&A, 454, 437 \\
Sanders J. S., Fabian A. C., 2007, MNRAS, 381, 1381 \\
Ulmer, M. P. et al. 1987, ApJ, 319, 118  \\
Van Speybroeck L. P., 1979 Vol. 0184, Einstein observatory (heao-b) mirror design and performance. pp 2–11 \\
Walker, S. A., Fabian, A. C., Russell, H. R., \& Sanders, J. S. 2014, MNRAS, 442, 2809\\
Zhuravleva I., Churazov E., Sunyaev R., Sazonov S., Allen S. W., Werner N., Simionescu A., Konami S., Ohashi T., 2013, MNRAS, 435, 3111 \\

\end{document}